\documentclass[conference]{IEEEtran}
\IEEEoverridecommandlockouts

\usepackage{amsmath,amssymb,amsfonts}
\usepackage{algorithmic}
\usepackage[T1]{fontenc}
\usepackage{textcomp}

\usepackage{xspace}
\usepackage{graphicx}
\usepackage{color}
\usepackage[sort,compress]{cite}
\usepackage{subcaption}
\definecolor{accessbl}{cmyk}{1, 0.4, 0.3, 0}
\DeclareCaptionFont{bl}{\color{accessbl}}
\usepackage{epsfig,endnotes,paralist,multirow}
\usepackage{wrapfig}
\usepackage{epsfig}
\usepackage[normalem]{ulem}
\usepackage{url}
\usepackage{soul}
\usepackage{multirow}
\usepackage{array}
\usepackage{bbm}

\def\includeComments{include}
\def\includ{include}
\def\comm[#1]{\ifx\includeComments\includ  \bl \texttt{\textbf{\textit{Note: #1}}} \par \fi}
\def\inlinecomm[#1]{\ifx\includeComments\includ  \textit{Note: #1} \fi}

\newcommand{\para}[1]{\smallskip \noindent {\bf #1}}
\newcommand{\softpara}[1]{\smallskip \noindent \underline{#1}}
\newcommand{\vsp}{\vspace{0.01in}}

\def\blackbox{\hfill {\vrule height6pt width6pt depth0pt}}

\newcounter{theorem}
\newtheorem{thm}{Theorem}

\newtheorem{corollary}{Corollary}

\newtheorem{defin}{Definition}
\newtheorem{ex}{Example}
\newtheorem{lem}{Lemma}
\newtheorem{ob}{Observation}

\newenvironment{thm-prf}{\vsp \begin{thm} \nopagebreak}{\end{thm}}

\newenvironment{lem-wo-prf-box}{\vsp \begin{lem} \nopagebreak}{\end{lem}}
\newenvironment{lem-prf}{\vsp \begin{lem} \nopagebreak}{\end{lem}}

\newenvironment{cor-prf}{\vsp \begin{corollary} \nopagebreak}{\end{corollary}}

\newcounter{packednmbr}

\newcommand{\EP}{EP\xspace}

\newcommand{\EPs}{EPs\xspace}
\newcommand{\dqc}{{\tt DQC}\xspace}

\newcommand{\E}{\mbox{${\cal E}$}\xspace}

\newcommand{\eps}{\mbox{EP}\xspace}
\newcommand{\epss}{\mbox{EPs}\xspace}

\newcommand{\eat}[1]{}

\let\emph\textit

\newcommand{\QAP}{\mbox{\tt QAP}\xspace}
\newcommand{\exQAPEta}{\mbox{\tt exQAP-$\eta$}\xspace}
\newcommand{\GreedyEta}{\mbox{\tt Greedy-$\eta$}\xspace}
\newcommand{\AnnealingEta}{\mbox{\tt Annealing-$\eta$}\xspace}

\newcommand{\PD}{\mbox{\tt DQC-PD}\xspace}

\newcommand{\NetworkC}{\mbox{\tt Network-C}\xspace}
\newcommand{\GreedyC}{\mbox{\tt Greedy-C}\xspace}
\newcommand{\SGDC}{\mbox{\tt SGD-C}\xspace}

\newcommand{\exQAPEtaT}{\mbox{\bfseries\boldmath exQAP-$\eta$}\xspace}
\newcommand{\GreedyEtaT}{\mbox{\bfseries\boldmath Greedy-$\eta$}\xspace}
\newcommand{\AnnealingEtaT}{\mbox{\bfseries\boldmath Annealing-$\eta$}\xspace}
\newcommand{\NetworkCT}{\mbox{\bf Network-C}\xspace}
\newcommand{\GreedyCT}{\mbox{\bf Greedy-C}\xspace}
\newcommand{\SGDCT}{\mbox{\bf SGD-C}\xspace}

\newcommand{\mpfont}{\scriptsize}

\ifx\noeditingmarks\undefined
    \newcommand{\MPworker}[2]{{\color{#1}\vrule\vrule}{\marginpar{\color{#1}\mpfont #2}}}
\else
    \newcommand{\MPworker}[2]{}
\fi

\def\BibTeX{{\rm B\kern-.05em{\sc i\kern-.025em b}\kern-.08em
    T\kern-.1667em\lower.7ex\hbox{E}\kern-.125emX}}

\begin{document}

\urldef\paperdoi\url{https://doi.org/10.1109/QCNC69040.2026.00043}
\title{Distributing Quantum Circuits Using Pre-Distributed Entanglement Pairs over Quantum Networks%
\thanks{Full author version revised after acceptance, including the cost-aware
entanglement consumption section omitted to meet the conference page limit.
The shortened conference version was published in
\emph{2026 International Conference on Quantum Communications, Networking, and Computing (QCNC)},
pp.~252--258. DOI: \paperdoi.
\par\textcopyright{} 2026 IEEE. Personal use of this material is permitted.
Permission from IEEE must be obtained for all other uses, in any current or future media,
including reprinting/republishing this material for advertising or promotional purposes,
creating new collective works, for resale or redistribution to servers or lists,
or reuse of any copyrighted component of this work in other works.}}

\author{
\IEEEauthorblockN{Yukun Yang, Ranjani G. Sundaram, Himanshu Gupta}
\IEEEauthorblockA{\textit{Department of Computer Science, Stony Brook University, Stony Brook, NY 11790, USA}}
}

\maketitle

\thispagestyle{plain}
\pagestyle{plain}

\begin{abstract}
Distributing quantum circuits across a network of quantum devices is a recent approach to improve the scalability of quantum computation. Entangled pairs of qubits distributed across network nodes are used as a communication resource for executing inter-node (remote) gates. However, establishing these entanglement pairs incurs significant latency, and doing so during the circuit's execution may lead to the decoherence of logical qubits. Instead, it may be beneficial to distribute certain entanglement pairs before the circuit is executed. Thus, we consider the problem of distributing a quantum circuit across a quantum network containing some pre-distributed entanglement pairs and seek to minimize the execution time of the circuit.

We address this problem in three steps. First, given the quantum network and (optionally) a probability distribution over training circuits, we determine a set of pre-distributed entanglement pairs under a specified budget. Second, we compute a qubit-to-memory allocation that minimizes the expected execution time given these pre-distributed resources. Finally, we identify any additional entanglement pairs required during execution, generate them concurrently, and execute the circuit gates. We evaluate our methods on a broad range of benchmark circuits.

\end{abstract}

\section{\bf Introduction}
\label{sec:intro}

The distribution of quantum circuits across a network of
quantum devices is a recent approach to improving the scalability of quantum computation. By connecting smaller quantum processors through quantum and classical communication channels, it becomes possible to perform computations (via remote gates) that would exceed the capacity of any single device. However, maintaining a quantum channel requires replenishing entangled pairs (\EPs) across adjacent network nodes. Establishing these entanglement pairs during the execution of a circuit can incur significant latency and lead to the decoherence of qubits. Thus, it is preferable to exploit existing \EPs that have been distributed before the execution of the circuit. Therefore, in this work, we distribute a given quantum circuit across a quantum network and seek to minimize circuit execution time by exploiting pre-distributed \EPs and generating required \EPs with minimum latency. 

In particular, before execution begins, we allocate a budget to actively generate and refresh a subset of \epss across the network. These pre-distributed \epss are chosen strategically, based on the network topology and (optionally) a probability distribution of anticipated circuits, so that frequently used or high-cost remote connections are established in advance. Unlike replenishment-based approaches, these \epss are one-time resources that are consumed during circuit execution, making their placement and distribution a critical design choice.

The distribution of quantum circuits then involves three stages: (i) generate pre-distributed \epss across the network under the given budget, (ii) map the qubits of the circuit to the qubit memories of the quantum network to minimize the expected execution time given the pre-distributed resources, and (iii) execute the circuit by generating any additional \epss concurrently and consuming these newly created pairs and existing ones to perform remote gate operations. Our proposed approaches are designed to maximize the utilization of pre-distributed \epss while minimizing the latency of additional entanglement generation during execution.

\para{Prior Work.}
The problem of distributing
quantum circuits in quantum networks has gained significant attention in recent years, resulting in the development
of efficient solutions tailored to various settings and objectives. Most prior work on distributing  quantum circuits has focused on the objective of minimizing the number of 
maximally-entangled pairs (\EPs) either by minimizing the number of 
cat-entanglements~\cite{Andres-MartinezH:19, g2021efficient} or the number of teleportations~\cite{zomorodi2018optimizing, sundaram2023distributing, Nikahd_2021}.
Several recent works~\cite{Caleffi,Ferrari-etal-2021, Yang2025Efficient,dqcqr, sundaram2025qcnc} have focused on the 
more significant objective of minimizing the total
latency incurred in generating the \EPs needed for executing the remote gates. However, all of them only start generating \EPs \emph{during} circuit execution and do not consider cases where pre-distributed \EPs are available for consumption in the network. In addition, Fan et al.~\cite{fan2023optimized,fan2024distribution} studied the efficient distribution and purification of GHZ and graph states in quantum networks. Although we also evaluate GHZ generation as a benchmark circuit, these works focus on state distribution, whereas we address distributed circuit execution with pre-distributed \epss.

\eat{Consider a setting wherein we are given a quantum network and a quantum circuit to be executed. In our formulation, the network is provisioned with a pre-distribution budget $K$ before execution begins. These \epss are treated as one-time resources that are consumed during the circuit and not replenished, in contrast to replenishment-based models. The central challenge is therefore to decide how to allocate this limited budget of \epss so that the subsequent qubit-to-memory mapping and circuit execution incur minimal latency. We design several algorithmic approaches to address this problem.}

\para{Our Work.}
In this paper, we formulate the problem of distributing a quantum circuit across a network with pre-distributed entanglement pairs (EPs) to minimize its execution time under network and decoherence constraints. Our contributions are threefold:

(i) \emph{Pre-distributed EPs allocation.} We design and analyze multiple algorithms for selecting where to place a limited pre-distribution budget, guided by training circuits and network topology.

(ii) \emph{Qubit-to-memory allocation.} We extend and adapt classical optimization techniques (including greedy heuristics, an extended quadratic assignment formulation, and simulated annealing) to assign circuit qubits to network memories given the pre-distributed EPs.

(iii) \emph{Execution with additional EPs.} We integrate these allocations into a full execution model that accounts for decoherence and on-demand generation, ensuring correctness while minimizing latency.

\section{\bf Background and Models}
\label{sec:background}

\para{Quantum Circuit Representation.}
We represent an \emph{abstract quantum circuit} $C$ over a set of qubits $\mathcal{Q} = \{q_1, q_2, \ldots\}$ as a sequence of gates $\langle g_1, g_2, \ldots \rangle$ where each $g_t$ is a binary CNOT gate or a unary gate (a universal set of gates).
We represent binary gates as triplets $(q_i, q_j, t)$ where $q_i$ and $q_j$ are the two operands and $t$ is the time instant of the gate in the circuit, and unary gates as pairs $(q_i, t)$ where $q_i$ is the operand and $t$ is the time instant. 

\para{Quantum  Network (QN).}
A quantum network is a network of quantum computers (QCs) represented as a connected graph with nodes as QCs and edges representing (quantum and classical) communication links. Each computer has a certain amount of quantum memory to store the data/circuit qubits.

\para{Distribution of Quantum Circuits over QN.} 
Given a quantum network (QN) and a quantum circuit, we seek to 
distribute and execute the given circuit over the network in such a way that the
execution incurs minimum time. 
Distribution of a circuit
over a network essentially entails two aspects: (i) distributing the
circuit qubits across the QCs/nodes of the QN, and (ii) executing the
``remote'' binary gates (i.e., gates with operands on different QCs) efficiently. The main overhead in distributed execution comes from these remote gates, which require \EPs to be available  across the involved nodes.

\para{Executing Remote Quantum Gates.}
To execute such remote gates, we need to bring all operands' values into a single QC via quantum communication. 
Since direct transmission of qubits
is subject to unrecoverable errors,
we consider the following ways to 
communicate qubits across network nodes.

\begin{figure}
\centering
\includegraphics[width=\linewidth]{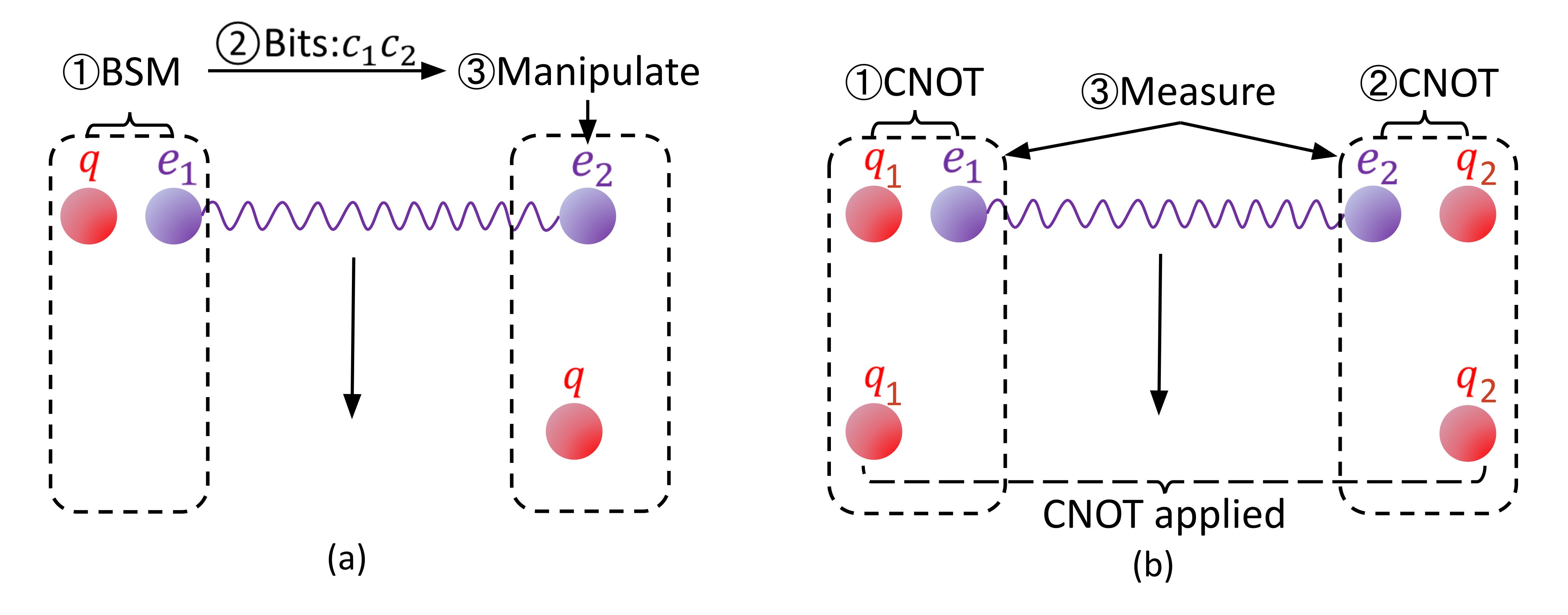}%
\caption{Quantum communication. (a) Teleportation and (b) Telegate. Dashed boxes are the network nodes; the initial (final) state of the qubits is at the top (bottom).}%
\label{fig:Communication}
\vspace{-0.2in}
\end{figure}

\softpara{Teleportation.}
An alternative approach to physically transmitting a qubit from a node
$A$ to $B$ is by \emph{teleportation}~\cite{Bennett+:93} 
which requires a priori distribution of an \EP over $A$ and $B$. 
See Fig.~\ref{fig:Communication}(a).
\eat{With an \EP distributed over $A$ and $B$, teleportation of a qubit 
state from $A$ to $B$ can be accomplished using classical 
communication and local gate operations while consuming/destroying 
the \EP. }
Teleportation moves the qubit state so that the binary gate can be executed locally at the destination QC.

\softpara{Telegates.} 
Another approach is the telegate protocol~\cite{Caleffi}, which executes 
a remote CNOT gate between operands at $A$ and $B$ using only local 
operations and classical communication. Telegates also require a 
priori \EP between $A$ and $B$. See Fig.~\ref{fig:Communication}(b).

\softpara{Teleportation vs.\ Telegates.}
Both communication protocols require a single
\EP. Teleportation transfers the qubit state to another node, while the telegate protocol executes the remote gate directly without qubit movement.
In this work, we focus on telegates for executing remote gates, as they align naturally with a static qubit-allocation function. The use of teleportations allows dynamic qubit allocations, but requires a more complex formulation, which we defer to future work.

\section{\bf Problem Formulation}
\label{sec:problem}

We start by defining some terms and models before moving on to the problem formulation.

\para{Qubit-Allocation Function.} Consider a quantum network with nodes (quantum computers) denoted as $\mathbb{P} = \{P_1, P_2, \dots, P_N\}$, where each node $P_i$ contains a set of qubit memories $M(P_i)$. For a quantum circuit $C$, let $Q(C)$ represent its set of qubits. To execute $C$ on the network, we define a \textit{qubit allocation function} $\eta: Q(C) \to \bigcup_i M(P_i)$, which assigns each circuit qubit to a distinct qubit memory at the network node. Specifically, $\eta$ must be one-to-one, ensuring that each qubit memory is assigned to at most one circuit qubit.

\softpara{Remote and Local Gates.} Given $\eta$, a gate $(q_i, q_j, t)$ in $C$ is classified as \textit{remote} if its operands $q_i$ and $q_j$ are assigned to different nodes, i.e., $\eta(q_i) \in M(P_u)$ and $\eta(q_j) \in M(P_v)$ with $u \neq v$. In contrast, a gate is \textit{local} if both operands reside at the same node.

\para{Pre-distributed \epss in Network.}
An important aspect in executing remote gates is how the entanglement resources are provisioned within the network. In our model, the network is initialized with a limited budget of \emph{pre-distributed \epss} before execution begins. These \epss are consumable one-time resources that can be consumed immediately during execution, thus reducing the need for online entanglement generation and minimizing overall execution latency. Since \epss are subject to decoherence, the network protocol refreshes the pre-distributed \eps pool during idle periods (when the network is not executing a circuit), so that the provisioned resources remain valid and available at the start of execution. During circuit execution, this background refresh is paused; only on-demand entanglement generation needed to cover unmet demand is performed. We assume that consuming the pre-distributed \epss (i.e., retrieving and using an already-established entangled pair during execution) incurs negligible online latency compared to on-demand entanglement generation and entanglement swapping.

\para{Coupling Graph and Swap Operations.} Each node $P_i$ has a \textit{coupling graph} $U_i$ that models the connectivity among its qubit memories. An edge $(m_a, m_b)$ in $U_i$ indicates adjacent qubits capable of direct gate execution. For non-adjacent qubits $m_c$ and $m_d$ in $P_i$, executing a gate requires swap operations along the shortest path between them in $U_i$. The number of swaps equals the path length in $U_i$.

\para{Static Qubit-Allocation Enforcement.} In general, the qubit-allocation function can change during circuit execution due to teleportations and/or swap operations. However, to simplify algorithm design and analysis, we enforce a static $\eta$ throughout execution. This implies:  
\begin{enumerate}
    \item \emph{Remote gates.} We execute remote gates exclusively via telegates, and do not teleport data qubits between nodes. Allowing teleportation-based routing would make $\eta$ time-varying and requires modeling dynamic mappings, which is outside our scope.
    \item \emph{Local gates with swaps.} When a local gate requires swaps, we swap qubits to bring them together, apply the gate, and then reverse the swaps to restore the original allocation.
\end{enumerate}

\para{Distributed Quantum Computation with Pre-distributed \epss (\PD) Problem.}
Given a quantum circuit $C$, a quantum network, and a pre-distribution budget 
$K$, the \PD problem is to execute $C$ across the network with minimum total execution time, subject to network constraints such as limited quantum memories at each node, entanglement generation resources, and decoherence. The \PD problem has three key components:
(i) determining how to allocate the limited pre-distribution budget across the network,
(ii) computing a static qubit-allocation function $\eta$ that jointly accounts for the pre-distributed \epss and on-demand entanglement requirements, and
(iii) devising an execution scheme for local and remote gates given all the above, so that the overall circuit execution time is minimized.

\softpara{Remarks and Computational Complexity.}
Our formulation differs fundamentally from previous work due to the presence of pre-distributed \epss. Instead of beginning directly with qubit allocation, we must first decide how to allocate a limited pre-distribution budget and then compute the qubit allocation function accordingly. This joint optimization of entanglement allocation and qubit placement leads to a combinatorial problem that is computationally intractable. Even when the set of pre-distributed \epss is fixed, determining the optimal qubit-to-memory assignment requires jointly reasoning about all pairwise gate interactions and network distances, a structure analogous to the classical quadratic assignment problem, which is NP-hard. Consequently, exact optimization is infeasible for realistic circuit and network sizes, motivating the development of efficient heuristic algorithms that approximate optimal allocation and execution decisions in the subsequent sections.

\section{\bf Related Works}
\label{sec:related}

\softpara{Minimizing the Number of EPs.} 
The \dqc problem is to distribute a given quantum circuit over a quantum network with some 
optimization objective. In contrast to the \PD problem, the \dqc 
problem ignores the coupling graphs within each 
computer~\cite{Andres-MartinezH:19, g2021efficient,Daei2020OptimizedQC,Nikahd_2021,dqc-gen,sundaram2023distributing}, and thus 
the needed swap operations. The optimization objective used in almost all 
of these works has been to minimize the communication 
cost, defined as the {\em number} of EPs used; a couple of recent works~\cite{MaoProbability,MaoQubitAllo} consider EPs with arbitrary (but fixed) cost and 
develop a simulated-annealing heuristic for the qubit-allocation part of the 
\dqc problem.

\softpara{Minimizing Circuit Execution Time.} 
The works close to ours~\cite{Caleffi, Ferrari-etal-2021, Yang2025Efficient, dqcqr, sundaram2025qcnc} address the
\dqc\ problem with minimizing circuit execution time, but all assume that the \EPs required are generated during circuit execution and do not consider pre-distributed \EPs. 
~\cite{Ferrari-etal-2021} focuses on estimating 
only the {\em worst-case} overhead in executing a circuit over a network,
by the worst-case linear network topology; the overhead considered is in terms of an increase in circuit ``layers'' and \EPs required to execute remote gates.
~\cite{Caleffi} focuses on the efficient execution of remote gates using telegates, {\em given} a qubit allocation (they consider qubit allocation to be out of the scope of their work), to minimize the number of circuit layers; they assume generation latency of each \EP to be uniform (a constant), and ignore decoherence constraints.
~\cite{Yang2025Efficient} considers the more general {\em multi-circuit} setting, where several circuits are executed concurrently over the same network; they design scheduling and resource allocation schemes to minimize the overall execution time across circuits.
~\cite{dqcqr,sundaram2025qcnc} are the works closest to ours, tackling the \dqc problem in three steps-allocating qubits to memories, selecting teleportations, telegates, and cat-entanglements for remote gate execution, and generating the necessary \EPs. The \PD problem is an extended formulation of their \dqc problem and considers pre-distributed \EPs.

\softpara{Pre-Distribution Framework.} 
Our problem formulation also differs fundamentally from the pre-distribution framework of Ghaderibaneh et al.~\cite{MG2022predistribution}. 
Their model selects super-links in the network to continuously replenish entanglement, thereby reducing the latency of future EP requests across arbitrary node pairs. 
In contrast, our goal is to execute a specific quantum circuit with the help of pre-distributed \EPs, which are provisioned before each execution and treated as one-time consumable resources. 
This setting does not model a continuous replenishment process during circuit execution; instead, resources are provisioned (and refreshed) between executions and then consumed during execution.
The central challenge then becomes how to allocate a limited pre-distribution budget so that the subsequent qubit allocation and circuit execution incur minimal latency.

\section{\bf High-Level Approach}
\label{sec:highlevel}

In the previous section, we formulated the \PD Problem as minimizing circuit execution time over a network given a limited pre-distribution budget. We use a three-step approach, consisting of allocating pre-distributed \epss, determining the qubit-allocation function, and finally determining the execution scheme.

\begin{enumerate}
\item
\emph{Allocate Pre-distributed \epss.} Given a quantum network, a budget $K$ of pre-distributed \epss, and (optionally) a distribution of training circuits, we determine how to allocate the \epss across memory pairs in the network to maximize their impact in reducing the overall execution time.

\item
\emph{Determine the Qubit-Allocation Function.} Given a quantum circuit, the network, and the chosen set of pre-distributed \epss, we compute the qubit-allocation function $\eta$ that maps circuit qubits to network memories in a way that minimizes the expected circuit execution time. Together with Step~1, this forms the primary focus of our work.

\item
\emph{Determine the Execution Scheme.} With the qubit-allocation function $\eta$ fixed and the set of pre-distributed \epss available, we devise an execution scheme that involves determining when and how to generate the additional \epss required to execute the remote gates. The goal is to minimize the total execution time.
\end{enumerate}

The final step largely follows prior work~\cite{dqcqr}: given a qubit allocation, we identify the remote gates, distinguish those already covered by pre-distributed \epss, and determine the set \E of additional \epss that must be generated during execution.

\para{Handling Decoherence.} Note that in the presence of decoherence, \EPs can decay both during and after their generation while waiting to be consumed. To mitigate this, we may need to partition \E into subsets of \EPs (called {\em batches}) with each batch generated independently, as generating the entire set concurrently could cause some \EPs to exceed the decoherence threshold $\tau$. We use the greedy algorithm from \cite{dqcqr} to construct a sequence of batches, each of which can be generated concurrently without violating the decoherence constraints.

The pre-distributed \epss may also decohere during execution. To mitigate decoherence, the network refreshes the pre-distributed EP pool between executions / during idle periods, so that provisioned EPs remain valid at the start of each execution. Together with the batching strategy for additional \epss, this jointly ensures that decoherence does not compromise the accuracy of execution.



\section{\bf Pre-distributed EPs Allocation}
\label{sec:preEP}

We now address the problem of determining how to allocate a limited pre-distribution budget across the network. This allocation step determines the matrix $C$ over memory pairs, which directly influences the subsequent qubit-allocation and execution steps. Recall that the pre-distributed EPs are prepared before execution and are treated as instantly consumable.

\para{Pre-distributed EPs Allocation Problem Formulation.} 
Given a quantum network, a budget $K$ of pre-distributed \epss, and (optionally) a probability distribution over circuits, we construct the following graphs and matrices.

\softpara{Circuit Graph $G_C$.} 
The circuit graph is an edge-weighted graph defined over the qubits $Q(C)$ of a given circuit $C$, 
with weight $w(q_i,q_j)$ on the edge $(q_i,q_j)$ equal to the number of binary gates between $q_i$ and $q_j$. 

\softpara{Network-Coupling Graph $G_N$.} 
The network-coupling graph is an edge-weighted graph defined over the set of qubit memory $\bigcup_i M(P_i)$ 
in the given quantum network with quantum computers $\{P_i\}$. 
For two memories $u,v$, the weight $w'(u,v)$ is defined as follows:
\begin{itemize}
\item If $u,v \in M(P_i)$ for some computer $P_i$, then $w'(u,v)$ equals the swap latency 
times twice the shortest path distance between $u$ and $v$ on the coupling graph of $P_i$. Note that this represents the time it takes to execute the gate over qubits in $u$ and $v$ using swap operations.
\item If $u \in M(P_i)$ and $v \in M(P_j)$ for $i \neq j$, then $w'(u,v)$ is the latency of 
independently generating an \eps between the network nodes $P_i$ and $P_j$. 
\end{itemize}

\softpara{\eps Allocation Graph $G_E$.}
The EP allocation graph is defined over the same set of memories, with weight $w''(u,v)$ equal to the 
number of pre-distributed \epss between memories $u$ and $v$. The budget constraint is 
\[
\sum_{u<v} w''(u,v)\, w'(u,v) \le K.
\]

\softpara{Qubit-Allocation Function $\eta$.} 
As introduced earlier in Sec.~\ref{sec:problem}, a qubit-allocation function $\eta : Q(C) \rightarrow \bigcup_i M(P_i)$ assigns each circuit qubit to a unique memory in the network.

\softpara{Pre-distributed EPs Allocation Problem Formulation.} 
Equivalently, let $A$ denote the adjacency matrix of the circuit graph $G_C$, 
$B$ denote the adjacency matrix of the network-coupling graph $G_N$, 
and $C$ denote the adjacency matrix of the \eps allocation graph $G_E$. 
For a given $\eta$, the execution cost of circuit $A$ is
\[
J(A,C,\eta) \;=\; 2\sum_{i<j} 
   \max\!\big(A_{ij} - C_{\eta(i),\eta(j)}, 0\big)\, B_{\eta(i),\eta(j)}.
\]
Thus, interactions covered by $C$ do not incur additional online latency, and only the unmet demand $\max\!\big(A_{ij}-C_{\eta(i),\eta(j)},0\big)$ contributes to the execution cost. For a distribution of representative circuits $\{A_r\}$ with probabilities $\{p_r\}$, 
the expected execution cost under allocation $C$ is
\[
\mathtt{COST}(C) \;=\; \sum_r p_r \min_\eta J(A_r, C, \eta).
\]
The pre-distributed \epss allocation problem is to select a feasible allocation matrix $C$ 
that minimizes $\mathtt{COST}(C)$ subject to the budget constraint above. 
Intuitively, a "good" allocation places \epss on memory pairs that are either structurally central in the network 
or highly demanded by the expected workload, thereby reducing the need for costly on-demand entanglement generation.

\para{Assumption: Negligible Latency to \emph{Consume} a Pre-distributed EP.}
Our execution-latency objective accounts for \emph{on-demand} entanglement generation as the dominant time component, and treats the pre-distributed EP as an already-established resource that can be consumed immediately when executing a remote operation.
Concretely, for a memory pair $(u,v)$, if the pre-distributed supply $C_{u,v}$ covers the circuit demand, then those covered remote interactions incur no additional online entanglement-generation latency.
Only the deficit must be generated during execution, which is captured by the term $\max\!\big(A_{ij}-C_{\eta(i),\eta(j)},0\big)$ in $J(A,C,\eta)$.
This matches our system model in which the network is initialized with a limited pre-distribution budget (before execution begins), and these EPs are consumable one-time resources that reduce the need for online generation.

In the following, we present \underline{three} algorithms for solving the pre-distributed \epss allocation problem: 
\NetworkC, \GreedyC, and \SGDC. 
They differ in their information requirements: 
\NetworkC\ uses only the network topology, while \GreedyC\ and \SGDC\ exploit additional knowledge of 
representative training circuits (or anticipated workload distributions). 

\para{Algorithm 1. Network-based EP Allocation (\NetworkCT).} 
The \NetworkC\ algorithm is an efficient topology-based heuristic that allocates pre-distributed \epss 
using only the structural properties of the network. It does not require workload or circuit information. 
The intuition is to favor memory pairs that are well-connected (high degree), close in distance, 
and uncongested. The steps are as follows:

\begin{enumerate}
\item 
\emph{Initialization.} Start with $C=0$ and set all link loads $L_{e}=0$ for edges $e$ on the network. 
Here $L_{e}$ denotes the number of pre-distributed \epss whose allocation paths traverse the edge $e$. 

\item 
\emph{Score Computation.} For each memory pair $(u,v)$, compute the score
\[
s_{u,v} = \frac{\deg(u)+\deg(v)}{B_{u,v}\,(1+C_{u,v})\,(1+L_{u,v})},
\]
where $\deg(\cdot)$ is the degree in $G_N$, $B_{u,v}$ is the distance of the network, 
$C_{u,v}$ is the current number of \epss allocated on $(u,v)$, 
and $L_{u,v}$ is the \emph{congestion of the path}, defined as the sum of $L_e$ over all edges $e$ 
on the shortest path between $u$ and $v$. 

\item 
\emph{Selection.} Identify the pair $(u,v)$ with the highest score $s_{u,v}$, 
assign one \eps\ to it, update $C_{u,v}$, and increase $L_e$ along the shortest path 
between $u$ and $v$. 

\item 
\emph{Iteration.} Repeat Steps 2–3 until the budget is exhausted. 
\end{enumerate}

By penalizing pairs whose shortest path is already heavily used, the congestion term $L_{u,v}$ ensures that allocations are spread more evenly across the network 
and avoids overloading specific links. 

\para{Algorithm 2. Greedy EP Allocation (\GreedyCT).} 
The \GreedyC\ algorithm assigns \epss\ to some top-ranked memory pairs in each round. Unlike \NetworkC, it explicitly leverages information from representative circuits (or anticipated workloads) to prioritize pairs with the highest unsatisfied demand. The steps are as follows:

\begin{enumerate}
\item 
\emph{Initialization.} Start with $C=0$ and compute the optimal qubit allocation function $\eta_r$ for each training circuit $A_r$\footnote{ For each training run of $C$, we fix a single qubit allocation algorithm (e.g., \GreedyEta) to serve as the $\eta$ oracle. The algorithms to determine $\eta$ are described in Sec.~\ref{sec:allocation}.}.

\item 
\emph{Score remote pairs.} For each remote pair $(u,v)$, compute its \emph{positive remainder mass}
\[
\begin{aligned}
ss_{u,v} &= \sum_r p_r \cdot
\mathbf{1}\!\big\{ A_r[i,j] > C_{u,v} \big\},\\
&\qquad (u,v) = (\eta_r(i),\eta_r(j)),
\end{aligned}
\]
which is the probability-weighted fraction of circuits where the demand on $(u,v)$ 
exceeds the current allocation $C_{u,v}$. 

\item 
\emph{Select top candidates.} Sort remote pairs by score, and choose the top-$k$ pairs. 
Filter by affordability with respect to the remaining budget. 

\item 
\emph{Commit and update.} Assign one \eps\ to each selected pair, update $C$, 
and refresh the optimal qubit allocation function $\eta_r$.

\item 
\emph{Iteration.} Repeat Steps 2–4 until the budget is exhausted. 
\end{enumerate}

\para{Algorithm 3. Subgradient Descent EP Allocation (\SGDCT).} 
The \SGDC\ algorithm formulates the allocation as a continuous optimization problem and applies the projected subgradient descent. It systematically exploits workload information to guide allocations. The steps are as follows:

\begin{enumerate}

\item 
\emph{Initialization.} Start with $C=0$ and a step size $\alpha$. Compute the optimal qubit allocation function $\eta_r$ for each training circuit $A_r$.

\item 
\emph{Subgradient update.} For each memory pair $(u,v)$, 
compute the subgradient using Danskin’s theorem\footnote{Danskin’s theorem states that if $V(x) = \min_y f(x,y)$, then a subgradient of $V$ with respect to $x$ is given by the partial derivative of $f(x,y)$ with respect to $x$, evaluated at any minimizer $y^*(x)$. In our case, $x=C$, $y=\eta$, and $f(C,\eta) = J(A,C,\eta)$.}:
\[
\begin{aligned}
G_{u,v} &= -B_{u,v} \sum_r p_r \cdot
\mathbf{1}\!\big\{ A_r[i,j] > C_{u,v} \big\},\\
&\qquad (u,v) = (\eta_r(i),\eta_r(j)).
\end{aligned}
\]
\item 
\emph{Projection.} Update $C \leftarrow C - \alpha G$ 
and project back onto the feasible budget set 
$\sum_{u<v} C_{u,v}\, B_{u,v} \le K$. Refresh the optimal qubit allocation function $\eta_r$.

\item 
\emph{Iteration and quantization.} Repeat Steps 2–3 until convergence, 
then round $C$ to integers and apply a top-up step to spend the remaining budget. 
\end{enumerate}

\section{\bf Qubit Allocation under Pre-Distributed EPs}
\label{sec:allocation}

Having determined the allocation matrix $C$ of pre-distributed \epss across the network, we next address the problem of \emph{qubit allocation}: mapping the qubits of a given circuit onto the network memories. The allocation must jointly exploit the pre-distributed \epss in $C$ and the structure of the circuit so as to minimize the overall execution cost.

\para{Qubit-Allocation Problem Formulation.} For a given circuit adjacency matrix $A$, network-coupling matrix $B$, and EP allocation matrix $C$, a qubit-allocation function $\eta : Q(C) \to \bigcup_i M(P_i)$ assigns each circuit qubit to a distinct network memory. The execution cost under $\eta$ is
\begin{align*}
J(A,C,\eta) \;=\; 
2\sum_{i<j} \max\!\big(A_{ij} - C_{\eta(i),\eta(j)},\,0 \big) \; B_{\eta(i),\eta(j)}.
\end{align*}
The qubit-allocation problem is to determine an assignment $\eta^\star$ that minimizes this cost:
\[
\eta^\star \;=\; \arg\min_{\eta} J(A,C,\eta).
\]
When $C \equiv 0$, the problem reduces to the classical minimum quadratic assignment problem (min-\QAP)~\cite{QAPHardness}, which is NP-Hard. Thus, the qubit-allocation problem remains computationally intractable. In the following, we present three algorithms to solve the qubit-allocation problem: 
\GreedyEta, \exQAPEta, and \AnnealingEta. They represent complementary strategies for the intractable quadratic assignment: \GreedyEta offers a lightweight heuristic with local refinement, \exQAPEta leverages approximation ideas from the maximum QAP to exploit structural information, and \AnnealingEta uses probabilistic search to escape local minima.

\para{Algorithm 4. Greedy Qubit Allocation (\GreedyEtaT).} 
The \GreedyEta\ algorithm ranks qubits by gate load, seeds the first assignment using the $C/(1{+}B)$ score, then assigns the remaining qubits by a greedy marginal-cost rule, and finally applies a 2-opt swap-based local search. The steps are as follows:

\begin{enumerate}
\item \emph{Qubit Ordering.} Order qubits by the total gate load (row sum of $A$), in non-increasing order: $q_1,q_2,\ldots,q_n$.

\item \emph{Seeding.} 
Compute the row score
\[
s(u) \;=\; \sum_{v} \frac{C_{u,v}}{\,1 + B_{u,v}\,}\,,
\]
and set $\eta(q_1) \gets \arg\max_{u} s(u)$.

\item \emph{Greedy Placement.} 
For $i=2,\ldots,n$, assign $\eta(q_i)$ to an unassigned memory $\bar m$ that minimizes the current objective increment
\[
\Delta J_i(\bar m)
\;=\;
\sum_{j \in \{1,\ldots,i-1\}}
\max\!\big(A_{ij}-C_{\bar m,\eta(j)},\,0\big)\; B_{\bar m,\eta(j)}.
\]

\item \emph{2-opt Local Search.} 
Repeatedly consider swaps of two qubits $(p,q)$ and accept the first swap that yields a negative cost change 
$\Delta J_{p,q}=J(A,C,\eta^{\text{swap}(p,q)})-J(A,C,\eta)$, 
until no improving swap exists.
\end{enumerate}

\para{Algorithm 5. Extended Quadratic Assignment (\exQAPEtaT).}
The min-\QAP\ problem is inapproximable within any constant factor~\cite{QAPHardness} (unless P=NP).  
In contrast, the max-\QAP\ admits a 4-approximation when one graph satisfies the triangle inequality~\cite{Esther}.  
Our approach adapts this idea by incorporating the pre-distributed \EPs into the objective, transforming the allocation into an extended max-\QAP instance that can be tackled with assignment and local refinement. The steps are as follows:

\begin{enumerate}
\item \emph{Initialization.}  
Following the AHS heuristic\footnote{Arkin–Hassin–Sviridenko (AHS) proposed an assignment-seeding method for max-QAP that combines structure and load information via a surrogate cost matrix, then solves it with Hungarian~\cite{Esther}. We adapt their idea to incorporate pre-distributed EPs through $C$.}, 
compute qubit gate loads $g(i)=\sum_j A_{ij}$.  
Compute a greedy max-weight matching $M$ on $B$ and set
\[
b(u)=
\begin{cases}
B_{u,\pi(u)}, & \text{if }(u,\pi(u))\in M,\\
0, & \text{otherwise,}
\end{cases}
\]
then form residual scores
\[
s(u)=\sum_v B_{uv}\,\max(\bar a - C_{uv},0),\quad
\bar a=\tfrac{1}{n(n-1)}\sum_{i\ne j}A_{ij}.
\]
Construct $F\in\mathbb{R}^{m\times n}$ with $F_{u,i}=b(u)\,g(i)+s(u)$ and solve a Hungarian assignment on $F$ to obtain the initial $\eta$.

\item \emph{Iterative Hungarian Refinement.}  
Given $\eta$, form
\[
G_{i,u}=2\sum_j \max(A_{ij}-C_{u,\eta(j)},0)\,B_{u,\eta(j)}.
\]
Solve Hungarian on $G$ to get $\eta'$. If $J(A,C,\eta')<J(A,C,\eta)$, set $\eta\leftarrow\eta'$ and repeat; otherwise stop.

\item \emph{2-opt Local Search.}  
Apply first-improvement 2-opt swaps, accepting any $(p,q)$ with
\[
\Delta J_{p,q}=J(A,C,\eta^{\text{swap}(p,q)})-J(A,C,\eta)<0,
\]
until no improving swap exists.
\end{enumerate}

\para{Algorithm 6. Simulated Annealing (\AnnealingEtaT).}
Simulated annealing refines qubit allocations by probabilistically accepting suboptimal swaps to escape local minima, balancing exploration and exploitation. The steps are as follows:

\begin{enumerate}
\item \emph{Initialization.}
Start with seed allocation (e.g., \GreedyEta\ or random).
Set the temperature $T=T_0$ and the cooling rate $\alpha\in(0,1)$.

\item \emph{Annealing Loop.}
For a fixed number of iterations per temperature:
\begin{itemize}
  \item Pick two qubits $(p,q)$ uniformly at random; let $\eta'$ be $\eta$ with their memories swapped.
  \item Compute $\Delta J = J(A,C,\eta') - J(A,C,\eta)$.
  \item If $\Delta J<0$ accept; else accept with probability $\exp(-\Delta J/T)$.
\end{itemize}
After each temperature stage, update $T \leftarrow \alpha T$ and continue until $T$ is small or the iteration budget is reached.

\item \emph{Best-so-far.}
Return the best allocation encountered during the run.
\end{enumerate}

\section{\bf Generalization to Cost-Aware Consumption of Pre-distributed \EPs}
\label{sec:gen}

\begin{figure*}[t]
    \centering
    \includegraphics[width=\textwidth]{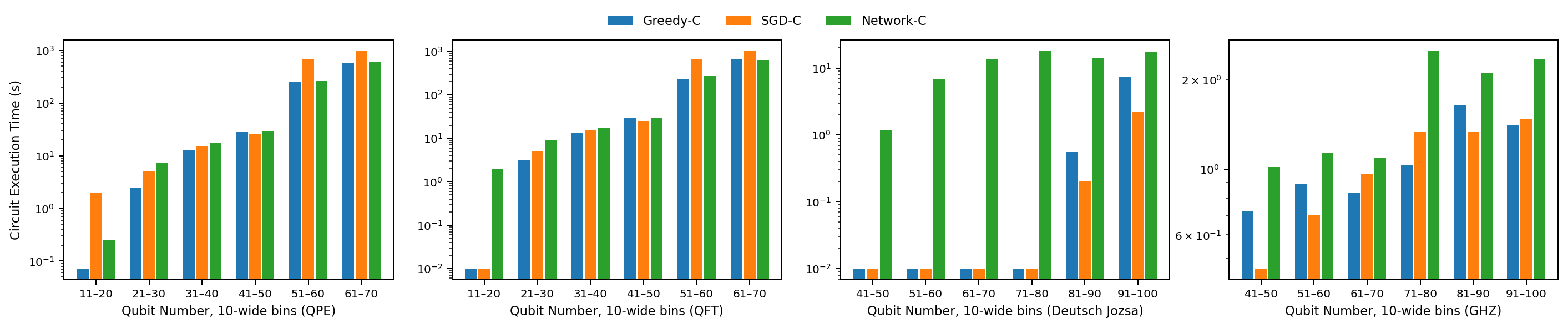}
    \caption{Execution time of four benchmark families (QPE, QFT, Deutsch Jozsa, GHZ) under three pre-distributed \EP trainers (\GreedyC, \SGDC, \NetworkC). Qubit allocation is fixed to \GreedyEta. Results are shown for 10-wide qubit bins; the $y$-axis is logarithmic.}
    \vspace*{-0.1in}
    \label{fig:C_function}
\end{figure*}

\begin{figure*}[t]
    \centering
    \includegraphics[width=\textwidth]{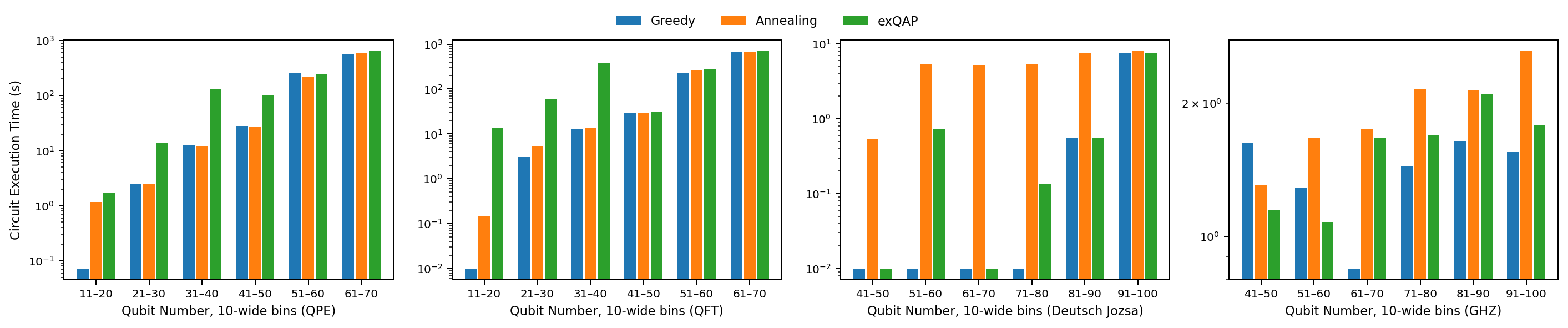}
    \caption{Execution time of four benchmark families under three qubit-allocation algorithms (\GreedyEta, \exQAPEta, \AnnealingEta). The pre-distributed \EP trainer is fixed to \GreedyC. Results are shown for 10-wide qubit bins; the $y$-axis is logarithmic.}
    \vspace*{-0.1in}
    \label{fig:eta_function}
\end{figure*}

\para{Motivation.}
Pre-distributed \EPs are shared latency-critical resources. In a multi-tenant quantum network, the operator may wish to discourage excessive consumption of pre-distributed \EPs, reserving them for situations where on-demand generation would be particularly costly. 
In practice, each pre-distributed \EP may incur a \emph{usage charge} based on its network distance or its freshness (e.g., remaining lifetime).
This motivates extending our formulation to account for the \emph{cost of consuming pre-distributed \EPs}, in addition to minimizing the execution latency.

\para{Price Model.}
We introduce a nonnegative \emph{price matrix} 
$\Pi = (\pi_{u,v})$ in memories, where $\pi_{u,v}$ denotes the usage 
charge per-\EP \ in pair $(u,v)$. A natural parameterization is:
\[
\pi_{u,v} = \beta_d \, B_{u,v} + \beta_f \, \varphi_{u,v},
\]
where $B$ is the network distance matrix, $\varphi_{u,v}\in[0,1]$ is a freshness 
score, and $\beta_d,\beta_f \ge 0$ are operator-defined weights. 

\para{Charge Constraint.}
For a circuit adjacency matrix $A$ and mapping $\eta$, the total consumption 
of pre-distributed \EPs is
\[
U(A,C,\eta) \;=\; \sum_{i<j} \min\!\big(A_{ij},\,C_{\eta(i),\eta(j)}\big),
\]
with an associated usage charge
\[
\Gamma(A,C,\eta;\Pi) \;=\; \sum_{i<j} \pi_{\eta(i),\eta(j)} 
   \cdot \min\!\big(A_{ij},\,C_{\eta(i),\eta(j)}\big).
\]

\para{Two Formulations.}
We consider two natural ways to incorporate usage charges:

\begin{itemize}
    \item {\em Hard-cap:} minimize execution latency subject to a usage budget:
    \[
    \min_{\eta} \; J(A,C,\eta)
    \quad \text{s.t.}\quad \Gamma(A,C,\eta;\Pi) \le \Phi.
    \]
    \item  {\em Soft-tradeoff:} include charges directly in the objective:
    \[
    \min_{\eta} \; J(A,C,\eta) \;+\; \lambda \,\Gamma(A,C,\eta;\Pi),
    \]
    where $\lambda \ge 0$ is a tunable tradeoff parameter.
\end{itemize}

\para{Objective Evaluation.}
For a fixed mapping $\eta$, abbreviate
\[
\begin{aligned}
d_{ij}^{\eta} &= \max\!\big(A_{ij}-C_{\eta(i),\eta(j)},0\big),\\
b_{ij}^{\eta} &= B_{\eta(i),\eta(j)},\qquad
p_{ij}^{\eta} = \pi_{\eta(i),\eta(j)}.
\end{aligned}
\]
Using $\min(x,y)=x-\max(x-y,0)$ and the definition of $J$, the
soft objective is exactly
\[
\begin{aligned}
F_{\lambda}(\eta)
&:= J(A,C,\eta)+\lambda\Gamma(A,C,\eta;\Pi)\\
&= \sum_{i<j}\big(2b_{ij}^{\eta}-\lambda p_{ij}^{\eta}\big)
   d_{ij}^{\eta}
   +\lambda\sum_{i<j}p_{ij}^{\eta}A_{ij}.
\end{aligned}
\]
The final sum generally depends on $\eta$, since prices vary across
memory pairs. It must therefore be retained when optimizing the mapping;
a shift of the distance matrix alone does not give an equivalent objective.

\para{Cost-Aware Search.}
The search frameworks in Sec.~\ref{sec:allocation} can be adapted to
evaluate the full objective $F_{\lambda}$. Greedy placement uses its
marginal increase over already assigned qubits. The 2-opt and annealing
methods use $F_{\lambda}(\eta')-F_{\lambda}(\eta)$ for their move tests,
while Hungarian-refinement proposals are accepted only if they improve
$F_{\lambda}$. These remain heuristic searches.
For the hard-cap formulation, candidate solutions must also satisfy
$\Gamma(A,C,\eta;\Pi)\le\Phi$ explicitly. Varying $\lambda$ can generate
candidate mappings, but does not guarantee feasibility or an optimal
constrained solution for this discrete problem.

\para{Discussion.}
Prices penalize mappings that consume expensive pre-distributed \EPs.
This extension retains the base model's consumption rule: for each mapped
pair, use $\min(A_{ij},C_{\eta(i),\eta(j)})$ pre-distributed \EPs and
generate the remaining demand on demand. Choosing to bypass an available
\EP\ would require an additional consumption decision.
The evaluation below uses $\lambda=0$; the cost-aware extension is not
evaluated experimentally.

\section{\bf Evaluation}
\label{sec:evaluation}

\para{Setup and Goals.}
We evaluate our approach in \emph{NetSquid}, a quantum-network simulator, using a telegate-based execution model~\cite{sundaram2025qcnc} for remote operations and the network/physics layer from Secs.~\ref{sec:preEP}--\ref{sec:allocation}. 
The pricing knob for pre-distributed usage is disabled ($\lambda=0$ in Sec.~\ref{sec:gen}), so the primary objective is to minimize execution time. 
We report results aggregated over benchmark circuits and over randomized network topologies, using common seeds across methods to enable paired comparisons.

\para{Benchmark Circuits.}
We evaluate four benchmark families from the Munich Quantum Toolkit~\cite{Quetschlich_2023}: Quantum Phase Estimation (QPE), Quantum Fourier Transform (QFT), Deutsch Jozsa, and GHZ-state generation (GHZ). 
Each family contains circuits of multiple sizes, which we group into inclusive qubit-number bins (e.g., $[11,20]$, $[21,30]$, \dots). 
For each bin, all circuits are used to train the pre-distributed \EPs allocation $C$. 
To assess performance, we then randomly sample $100$ circuits from the same bin as test instances. 
Each sampled circuit is executed with the trained $C$ and the chosen qubit-allocation method $\eta$, and their execution times are summed together. 
The identical set of $100$ test circuits is used across all methods to enable paired comparisons, and when $C$ is trained, the inner placement oracle $\min_{\eta}$ is instantiated with the \emph{same} $\eta$-method used at the test time (oracle-aligned training). 
This evaluation protocol allows us to compare fairly among $C$-trainers, $\eta$-methods, and their joint combinations.

\para{Generating Random Networks.}
The network comprises $10$ compute nodes, each with $10$ quantum memories. We use a network spread over an area of $100 km \times 100 km$\footnote{The 
$100 km \times 100 km$ area is a modeling choice to set a representative spread of inter-QPU distances; under uniform rescaling of all link lengths, the cost matrix $B$ (and budget $K$ if scaled accordingly) rescales by a constant factor, so the allocation decisions are driven primarily by relative inter-QPU costs rather than the absolute geographic scale.}. We use the Waxman model~\cite{waxman} to create Internet topologies, distribute nodes, and create links\footnote{Inter-QPU distances parameterize $B$ (inter-node EP cost), while intra-QPU connectivity affects local compilation overhead (e.g., swap operations).}. 

\softpara{Network Parameters.} We use values of network parameters similar to those used in~\cite{Ghaderibaneh2022}. In particular, we set the atomic-BSM probability of success and latency to be 0.4 and 10$\mu$ seconds and the optical-BSM probability of success to be 0.3. We use atom-photon generation times and probability of success as 50$\mu$ sec and 0.33, and the decoherence threshold of 1 second.

\softpara{Budget $K$.}
For each randomized network instance, we set the pre-distribution budget $K$ as
\[
K \;=\; \frac{N}{2}\cdot \overline{d}_{\mathrm{remote}},
\]
where $N$ is the total number of memories in the network and $\overline{d}_{\mathrm{remote}}$ is the average
shortest-path distance computed over unordered memory pairs that lie on different compute nodes.
This fixes a concrete, reproducible budget per network instance and scales with the network size and geometry.

\begin{figure*}[t]
    \centering
    \includegraphics[width=\textwidth]{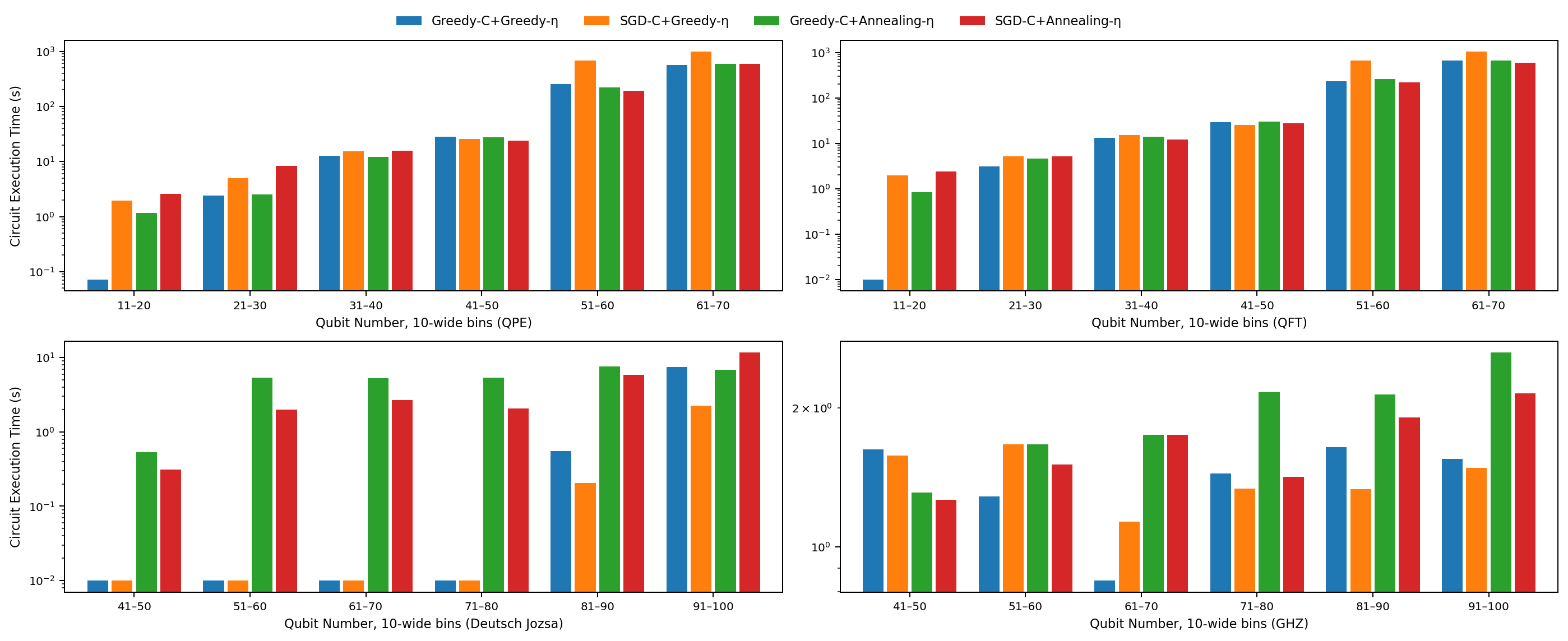}
    \caption{Execution time of four benchmark families under joint combinations of \EP trainers and qubit-allocation methods. We report oracle-aligned training, i.e., $C$ is trained using the same $\eta$-method applied at test time. Results are shown for 10-wide qubit bins; the $y$-axis is logarithmic.}
    \vspace*{-0.1in}
    \label{fig:C_eta_function}
\end{figure*}

\para{Algorithms Compared.}
We evaluate three types of comparison.
\begin{enumerate}
\item 
We compare the three algorithms for training pre-distributed \EPs from Sec.~\ref{sec:preEP}: \GreedyC, \SGDC, and \NetworkC, while fixing the qubit allocation to \GreedyEta\ (Fig.~\ref{fig:C_function}).
        
\item 
We compare the three algorithms for the allocation of qubits in Sect.~\ref{sec:allocation}: \GreedyEta, \AnnealingEta and \exQAPEta---while fixing $C$ to be trained by \GreedyC \ (Fig.~\ref{fig:eta_function}).

\item 
We evaluate \emph{oracle-aligned combinations}, where $C$ is trained using the same $\eta$-method that is applied at the test time, thus capturing the joint effect of $C$-training and $\eta$-allocation. 
In particular, we consider the $2\times 2$ set of combinations obtained from $\{\GreedyC, \SGDC\}$ crossed with $\{\GreedyEta, \AnnealingEta\}$ (Fig.~\ref{fig:C_eta_function}).
\end{enumerate}

\para{Evaluation Results.}
We evaluate the algorithms over the circuits and networks described above. Note that the $y$-axis is logarithmic and tiny values are clipped at the lower bound of the axis. We observe the following:

\begin{enumerate}
\item In the smallest QFT bin and the first four DJ bins in Fig.~\ref{fig:C_function}, \GreedyC, and \SGDC cover essentially all remote gates within the budget $K$, yielding near-zero execution time (clipped at the lower bound of the log scale). In other figures there are similar phenomena.

\item In Fig.~\ref{fig:C_function}, \GreedyC performs best on QPE and QFT, while \SGDC is stronger on Deutsch Jozsa; GHZ shows mixed behavior. The structure-only \NetworkC trainer underperforms the circuit-aware trainers.

\item In Fig.~\ref{fig:eta_function}, \GreedyEta performs best among the three algorithms for qubit allocation.

\item In Fig.~\ref{fig:C_eta_function}, it shows a trend similar to that in Fig.~\ref{fig:C_function}. \GreedyC +\GreedyEta performs best in QPE and QFT, while \SGDC+\GreedyEta is stronger in Deutsch Jozsa; GHZ shows mixed behavior.
\end{enumerate}

\softpara{Why Benchmark Families Differ.}
The circuit families induce different two-qubit demand profiles. QFT and QPE involve dense long-range couplings, so even small misplacements in $\eta$ leave substantial uncovered demand, and execution time grows rapidly with $n$. Deutsch–Jozsa has a much fewer effective demand; with a moderate budget $K$, most remote gates can be eliminated, keeping times close to the log-scale floor and reducing differences between methods. GHZ circuits form a chain-like pattern whose cost depends strongly on where the chain anchors in the network; as a result, assignment methods trade off differently, producing mixed winners across bins.

\para{Takeaway.}
Circuit-aware training of pre-distributed \EPs (\GreedyC, \SGDC) together with effective qubit placement (\GreedyEta) consistently achieves the lowest execution times. 
This underscores that optimizing both \EP allocation and qubit placement in tandem is crucial to minimize execution latency in distributed quantum computation.

\section{\bf Conclusions}
\label{sec:conc}

In this work, we addressed the \PD problem: executing a quantum circuit across a network with the aid of pre-distributed entanglement pairs. We developed and evaluated algorithms for both pre-distributed EP allocation and qubit-to-memory assignment, showing that their joint optimization is crucial for minimizing execution latency. Looking ahead, several extensions are natural. One is to incorporate operations based on teleportation and cat-entanglement into our framework. Another is to extend our algorithms to the online setting, where circuits arrive dynamically and EPs must be replenished intelligently. Addressing these challenges will further advance the goal of scalable and efficient distributed quantum computation.

\bibliographystyle{IEEEtran}
\bibliography{IEEEabrv,all_shorten}

\end{document}